\documentclass{sig-alternate}

\usepackage{cmap}
\usepackage[utf8]{inputenc}
\usepackage{microtype}
\usepackage{paralist}
\usepackage{xcolor}
\usepackage{balance}

\usepackage[hyphens]{url}
\usepackage{hyperref}

\usepackage{listings}
\usepackage{color}

\definecolor{lightgray}{rgb}{.9,.9,.9}
\definecolor{darkgray}{rgb}{.4,.4,.4}
\definecolor{purple}{rgb}{0.65, 0.12, 0.82}
\lstdefinelanguage{json}{
  basicstyle=\small\ttfamily,
  keywords={boolean, integer, string, uri, serialization\_format, object, policy\_object,
		json\_schema\_object, request\_object, response\_object, auth\_object, 
		interaction\_object, extractor\_object, null},
  keywordstyle=\color{purple}\bfseries,
  identifierstyle=\color{black},
  sensitive=true,
  stringstyle=\color{black}\ttfamily,
  morestring=[b]",
  showstringspaces=false,
  showspaces=false,
	tabsize=2
}
\lstdefinelanguage{python}{
  basicstyle=\small\ttfamily,
  keywords={def, import, for, in, print},
  keywordstyle=\color{purple}\bfseries,
  identifierstyle=\color{black},
  sensitive=true,
  stringstyle=\color{black}\ttfamily,
  morestring=[b]",
  morecomment=[l]{\#},
  commentstyle=\color{blue}\rmfamily,
  showstringspaces=false,
  showspaces=false,
	tabsize=2
}

\begin{document}

\toappear{Copyright is held by the author/owner(s).\par
\emph{WWW 2012 Developer Track}, April 18--20, 2012, Lyon, France.}

\title{API Blender: A Uniform Interface to Social Platform APIs}

\newcommand{\georges}[1]{[\textcolor{red}{\textbf{Georges:} #1}]}
\newcommand{\pierre}[1]{[\textcolor{green}{\textbf{Pierre:} #1}]}

\numberofauthors{2}
\author{
\alignauthor
Georges~Gouriten\\
       \affaddr{Institut Télécom; Télécom Paristech; CNRS LTCI}\\
       \affaddr{46, rue Barrault, Paris, France}\\
       \email{georges.gouriten@telecom-paristech.fr}
\alignauthor
Pierre~Senellart\\
       \affaddr{Institut Télécom; Télécom Paristech; CNRS LTCI}\\
       \affaddr{46, rue Barrault, Paris, France}\\
       \email{pierre.senellart@telecom-paristech.fr}
}
\additionalauthors{}

\newcommand{\system}{\textsc{API Blender}}

\maketitle

\begin{abstract}
With the growing success of the social Web, most Web developers have to
interact with at least one social Web platform, which implies
studying the related API specifications. These are often only informally
described, may contain errors, lack harmonization, and generally speaking
make the developer's work difficult. Most attempts to solve this problem,
proposing formal description languages for Web service APIs, have had
limited success outside of B2B applications; we believe it is due to
their top-down nature. In addition, a programmer dealing with one or
several of these APIs has to deal with a number of related tasks such as
data integration, requests chaining, or policy management, that are
cumbersome to implement. Inspired by the SPORE project, we present
\system{}, an open-source solution to describe, interact with, and
integrate the most common social Web APIs. In this perspective,
we first introduce two new lightweight description formats for requests
and services and demonstrate their relevance with respect to current 
platform APIs. We present our Python implementation of \system{}
and its features regarding authentication, policy management and
multi-platform data integration.
\end{abstract}

\category{H.3.5}{Information Storage and Retrieval}{Online Information Services}[Web-based services]
\category{D.3.2}{Programming Languages}{Language Classifications}[Python]

\terms{Design, Standardization}

\keywords{social Web, API, REST, data integration}

\clubpenalty=10000
\widowpenalty=10000

\section{introduction}
Interacting with platforms like Facebook,
Youtube, Twitter, Flickr, or Google+
becomes an important part of many software projects, whether it is for
authentication purposes, to collect information about a user, to present
mash-ups of popular social Web data, or for a myriad of other reasons.
Our perspective comes from the need of \emph{archiving} important social
data for preservation purposes.\footnote{ARCOMEM project,
\url{http://www.arcomem.eu/}} Regular Web
archives, such as those built by the Internet
Archive\footnote{\url{http://www.archive.org/}},
often include content from or pointers to social Web platforms but do not benefit 
from API data. As a consequence, the archives are either partial -- Facebook disallows generic crawling of its public
pages -- or lack some extra information that the API can provide, for
instance extracted entities on Twitter. Designing an \emph{archival
crawler} for the social Web requires interfacing with the multiple social
Web APIs, as well as respecting the \emph{policies} imposed by these
services, such as limiting the number of requests per hour.

Many projects thus involve numerous interactions with various social
platforms, sometimes with complex logics such as getting
the social graph till the third rank of users having mentioned a specific keyword.
Understanding the related API specifications can be challenging.
There is no \emph{de facto} standard to describe them and they can
contain mistakes or approximations. There is no clear specification, for
instance, of how many requests per hour are allowed on the Twitter search API.
For the most popular platforms, specific language libraries sometimes exist
but they often require the same learning phase.

Having a unified description of the different social Web APIs is a technical
 challenge. An early step was taken with WSDL~\cite{WSDL}, a Web Services
Description Language standardized by the W3C.
WSDL has been heavily used in the industry and is at the core of many
service-oriented software projects~\cite{erl:service}.
However, most popular social platforms including Facebook, Twitter, or
Google+ and many other Web
services are not currently offering any WSDL description of their
API and do not seem to have any plans to do so.
The reasons are manifold: WSDL-based services are often considered heavy
machinery for such lightweight
interfaces~\cite{DBLP:journals/debu/Alonso02}, WSDL has
historically focused on SOAP message exchanges rather than on RESTful
APIs though it can now express both \cite{mandel:wsdl}, WSDL has no support for important API
metadata such as policy management or the description of a sequence of
service calls\footnote{BPEL~\cite{BPEL} is typically used in B2B projects that need service
orchestration, but leads to even heavier machinery.}.
In reaction to WSDL, some other approaches to Web services
description have been proposed, a prime example being WADL~\cite{WADL}
but they have not met with more success on popular social Web platforms.

Another perspective is necessary.
Spring Social\footnote{\url{http://www.springsource.org/spring-social}} is a Java framework
to interact with the different social platforms. 
We believe this bottom-up approach is a very promising way to make the developers' work simpler.
Spring Social implements a number of useful functionalities
(authentication, uniform interface to some of the API types, etc.) but
does not fulfill our requirements. On the one hand, some important
features, especially for archival crawling, are not considered, such as
limits on the number of requests. On the other hand, using Spring Social
requires understanding
an important amount of code before being able to interact with a social 
platform. To give an order of magnitude of the size of the software, 
the core v1.02 contains 405 files, without implementing any Web
API.\footnote{\url{http://s3.amazonaws.com/dist.springframework.org/}}
With \system, we aim at more simplicity and flexibility, as highlighted by
the example of use we give in Section~\ref{sec:implementation}.

Our main source of inspiration has been the SPORE project~\cite{SPORE}. 
It consists in a simple implementation-agnostic JSON format allowing to describe 
Web APIs designed according to the REST principles. 
The project has been started recently and is still under development.

With \system, we extend SPORE with the following contributions:
\begin{compactenum}
\item{two simple description formats at the API and request levels,
 adapted to social platforms, sorting SPORE out and complementing it;}
\item{an open Python implementation, allowing to easily integrate various platforms;}
\item{the following features, some of them left out of
existing tools or libraries: authentication, server policy management,
multi-platform data integration, and request chaining.}
\end{compactenum}

We designed \system\ inspired by what we observed on five prominent social platforms
we identified: Twitter, Facebook, Google+, FlickR and Youtube. However, we
strove at keeping a high flexibility so that it can be extended to many other 
Web APIs.

Our article is organized as follows. In Section~\ref{sec:format}, we
present descriptions formats and
discuss their relevance to social platforms.  We then detail in
Section~\ref{sec:implementation} our implementation in Python and its features.

\section{Description formats}
\label{sec:format}
A Web API consists in a set of HTTP request messages associated to responses, sent
to a specific HTTP server having its own rules.
Note that Twitter has different APIs corresponding to different
hosts: for instance,
\url{api.twitter.com:80} or \url{search.twitter.com:80}.\footnote{\url{https://dev.twitter.com/docs/history-rest-search-api}}
We describe a Web API with several objects that allow to describe the server
and its rules (with respect to access policies) as well as the interactions it offers.
We find JSON~\cite{JSON} light and readable and have chosen to use it
as a serialization. In what follows, we tried using
straightforward names and self-explaining conventions to define the different
elements.

\paragraph*{Server description format}
We have extended SPORE with a consistent oriented-object approach,
as well as the addition of authentication and policy
usually required to interact with social platform Web APIs. 

\begin{lstlisting}[language=json, title={Server Object}]
"name": string,
"host": string,
"port": integer, 
"authentication": auth_object, 
"policy": policy_object, 
"interactions": [interaction_object] 
\end{lstlisting}
Port, policy, and authentication are
optional. The port defaults to 80.

Two authentication protocols are supported at the moment, one based on a
unique authentication URL with parameters and the other on the three-legged 
OAuth2~\cite{OAUTH2}.

\begin{lstlisting}[language=json, title={Simple Authentication Object}]
"request_token_url": uri,
"url_parameters": object
\end{lstlisting}
By simple authentication, we mean
authentication with parameters such as API key
or login and password passed to a unique URL so as to receive the
authentication token.

\begin{lstlisting}[language=json, title={OAuth2 Authentication Object}]
"consumer_key": string,
"consumer_secret": string,
"request_token_url": uri,
"access_token_url": uri,
"authorize_url": uri
\end{lstlisting}
Many social platforms (e.g., Twitter, Facebook, Google+) accept OAuth2 
authentication.

\begin{lstlisting}[language=json, title={Policy Object}]
"requests_per_hour": integer,
"too_many_calls_response_code": integer,
"too_many_calls_waiting_seconds": integer
\end{lstlisting}
An overload can be detected by counting the requests or receiving a 
too-many-calls response.
In the latter case, \system\ will snooze for the specified amount of time before testing if the counter has been reset.

\paragraph*{Interaction description format}
An interaction is a class of HTTP requests with a common root path and their associated responses.
Here also we extended SPORE and added the response object.

\begin{lstlisting}[language=json, title={Interaction Object}]
"name": string, 
"description": string,
"request": request_object,
"response": response_object
\end{lstlisting}
The description is optional.

\begin{lstlisting}[language=json, title={Request Object}]
"root_path": string, 
"method": string 
"raw_content": string 
"url_parameters": [
	[	string, # key, e.g., "id"
		string, # type,	e.g., "integer"
		boolean # is it an optional parameter?
		object  # the default value, it can be null ]
]
\end{lstlisting}
The method has to be GET, PUT, POST or DELETE. Providing raw content
is optional and useful only for PUT and POST methods. If a default value is
set on a URL parameter, it will be automatically passed with the default value
unless it is explicitly set as null. 
This feature can be useful in many case such as requesting 
a default value of 100 responses per pages for a full-text query on Twitter
search API.

\begin{lstlisting}[language=json, title={Response Object}]
"expected_status_code": integer,
"serialization_format": serialization_format,
"expected_schema": json_schema_object,
"integration": extractor_object
\end{lstlisting}
The expected code is optional and defaults to 200.
The serialization format has to be JSON or XML at the moment.
The expected schema of the response is optional and can be defined as a
JSON schema~\cite{JSON_SCHEMA}. At the moment, we define a simple extractor
that allows a mapping between a unified model and response fields. 
We use `.' as a path separator. For instance, we could have 
\lstinline[language=json]{"post.content": "post_data.text"}
if our integrated model was {\small\verb+{"post": {"content": string}}+} with
a response model of {\small\verb+{"post_data": {"text": string}}+}.
With a careful normalization model (for instance using concepts of an
ontology), this allows to integrate data coming
from different platforms. As an extension, this semantic model could also
be used to describe the inputs of services, a first step towards semantic
service orchestration.

\section{The Python implementation}
\label{sec:implementation}
Python is becoming increasingly popular among developers.
On the social coding platform GitHub, it is ranked third.\footnote{After JavaScript and Ruby, \url{https://github.com/languages}}
We find Python to be simple, flexible, and to have many useful libraries.
We have chosen to implementation \system\ in this language. 
\system\ is available online at \url{https://github.com/netiru/apiblender}.

\paragraph*{Structure}
The module structure offered by Python allows us to adopt the following light
structure.

\begin{lstlisting}[language=json, title={\system\ package}]
main.py 					 Controller
server.py					 Server and interactions
policy.py 				 Policy management
auth.py 					 Authentification management
config/ 					 JSON configuration files
--general.json 		 General config
--apis/						 API config files
\end{lstlisting}
We found it convenient to have one file per API server where we gather the 
descriptions for the server and its interactions. Currently, the API Blender
supports the two Twitter APIs (generic and search), Facebook, Google+,
FlickR and Youtube.
\paragraph*{Features}
\system\ implements several precious feature. It supports \emph{the two main authentication
types}: using a single URL with parameters and OAuth2~\cite{OAUTH2} thanks to Python OAuth2\footnote{Created and
maintained by SimpleGeo Inc. \url{https://github.com/simplegeo/python-oauth2}}.

\system\ also \emph{ensures respect of the server policy}; when the hourly limit is reached or when a 
too-many-calls response is identified, the policy manager 
will stop for some time and periodically test if the counter has been reset.
\emph{Error handling} is taken into consideration too, whether it regards a non-conforming
configuration file or an unexpected response.
Finally, \system\ gives the possibility to extract and normalize elements from responses.
This feature supports simple field extraction and standardization at the
moment
but the same process will be possible with arbitrary subtree
transformations in the near future.
\paragraph*{Request chaining}
The open nature of \system\ combined to the flexibility of Python can fill many needs.
Request chaining becomes very simple with Python and complex interactions can
become easy-to-maintain Python libraries.
We illustrate this with the following example on two Twitter APIs. The
program below retrieves the last three pages of tweets containing the keywords
``good spirit'' then fetches the local social network (followers and
followees) of the authors of the tweets.
\begin{lstlisting}[language=python, title={Example of request chaining with Python}]
import apiblender

blender = apiblender.Blender()

# Retrieving 3 pages of result
blender.load_server("twitter-search")
blender.load_interaction("search")
users = set()
for p in range(1,3): 
    blender.set_parameters({"q": "good spirit", 
                            "page": p})
    response = blender.blend()
    ts=response["prepared_content"]["results"]:

    for twitt in ts
        users.add(twitt["from_user"])

# Retrieving followers / followees for each user
blender.load_server("twitter-generic")
for user in users:
    blender.load_interaction("followers")
    blender.set_parameters({"screen_name":user})
    followers = blender.blend()

    blender.load_interaction("followees")
    blender.set_parameters({"screen_name":user})
    followees = blender.blend()

    # Printing everything
    print("User Name: %s" % user)
    print("\tFollowers: %s" % \
            followers["prepared_content"])
    print("\tFollowees: %s" % \
            followees["prepared_content"])
\end{lstlisting}

\pagebreak
\section{Conclusions}
\system\ has been designed in the context of the ARCOMEM
project on social Web archiving, and is 
put to use in this project to crawl and integrate data from various
social Web platforms. We have found its flexibility useful in the light
of the dynamicity of social Web platforms and managed to conveniently
integrate the five platforms currently supported: Twitter,
Facebook, FlickR, Google+, and Youtube.
It is of potential use in any application that
needs to access similar REST-inspired Web APIs and to export responses 
in a common schema. The source code being available on GitHub, we hope to
solicit contributions, either in
the form of extensions of the base functionalities, or in that of API
descriptions. For future work, we see many promising opportunities such as:
\begin{compactenum}
\item smarter processing of responses, making use of the semantics of the
services described, in the spirit of the semantic description of Web
services à la OWL-S~\cite{OWL-S};
\item developing more standard request chaining libraries;
\item a possible integration of the different input schemas;
\item more research for a smarter snooze management;
\item distributing requests across different servers.
\end{compactenum}
They all require to be very conscious of the existing trade-off between completeness and
flexibility.

\section{Acknowledgments}

The described work was funded by the European Union Seventh Framework
Programme (FP7/2007--2013) under grant agreement 270239 (ARCOMEM).

\balance
\bibliographystyle{abbrv} 
\bibliography{main}

\end{document}